\documentclass{article}
\usepackage{spconf,amsmath,graphicx,tikz, makecell, xcolor, caption, setspace}
\usepackage[inkscapelatex=false]{svg}
\usepackage{cite}

\usepackage{enumitem}
\setlist{nosep, leftmargin=14pt}

\usepackage{mwe} 

\title{Towards Patient-Specific Surgical Planning for Bicuspid Aortic Valve Repair: Fully Automated Segmentation of the Aortic Valve in 4D CT}
\name{
    \begin{minipage}{\textwidth}
        \centering
            Zaiyang Guo$^{\star}$, Ningjun J Dong$^{\star}$, Harold Litt$^{\star}$, Natalie Yushkevich$^{\star}$, Melanie Freas$^{\star}$, Jessica Nunez$^{\star}$, Victor Ferrari$^{\star}$, Jilei Hao$^{\star}$, Shir Goldfinger$^{\star}$, Matthew A. Jolley$^{\dagger}$, Joseph Bavaria$^{\S}$, Nimesh Desai$^{\star}$, Alison M. Pouch$^{\star}$
    \end{minipage}
}
\address{
    \begin{minipage}{\textwidth}
        \centering
        $^{\star}$ University of Pennsylvania, Philadelphia, PA, USA\\
        $^{\dagger}$Children's Hospital of Philadelphia, Philadelphia, PA, USA \\
        $^{\S}$Thomas Jefferson University, Philadelphia, PA, USA
    \end{minipage}
}

\begin{document}

%
\maketitle

\section{Abstract}
The bicuspid aortic valve (BAV) is the most prevalent congenital heart defect and may require surgery for complications such as stenosis, regurgitation, and aortopathy. BAV repair surgery is effective but challenging due to the heterogeneity of BAV morphology. Multiple imaging modalities can be employed to assist the quantitative assessment of BAVs for surgical planning.  Contrast-enhanced 4D computed tomography (CT) produces volumetric temporal sequences with excellent contrast and spatial resolution. Segmentation of the aortic cusps and root in these images is an essential step in creating patient-specific models for visualization and quantification. While deep learning-based methods are capable of fully automated segmentation, no BAV-specific model exists. Among valve segmentation studies, there has been limited quantitative assessment of the clinical usability of the segmentation results. In this work, we developed a fully automated multi-label BAV segmentation pipeline based on nnU-Net. The predicted segmentations were used to carry out surgically relevant morphological measurements including geometric cusp height, commissural angle and annulus diameter, and the results were compared against manual segmentation. Automated segmentation achieved average Dice scores of over 0.7 and symmetric mean distance below 0.7 mm for all three aortic cusps and the root wall. Clinically relevant benchmarks showed good consistency between manual and predicted segmentations. Overall, fully automated BAV segmentation of 3D frames in 4D CT can produce clinically usable measurements for surgical risk stratification, but the temporal consistency of segmentations needs to be improved.

\section{Introduction}
The bicuspid aortic valve (BAV) is the most common congenital heart defect with a prevalence of 0.5\% to 2\% \cite{ward_clinical_2000}. Among several common BAV complications, aortic regurgitation (valve leakage) necessitates surgical intervention when severe. Conventionally, the native valve can be surgically replaced with a mechanical or bioprosthetic valve. However, mechanical valves require lifelong anti-coagulation, which can impact quality of life and increase bleeding risk \cite{svensson_long-term_2014}. Bioprosthetic valves have limited durability and require re-intervention, making them less suitable for younger patients \cite{forcillo_perimount_2014}. In recent years, BAV repair surgery has emerged as an alternative to replacement long-term durability reaching 15 to 20 years, in valves that are well-suited for this option \cite{ehrlich_state--art_2020}.

The main challenge of BAV repair lies in its complexity. BAVs vary significantly in morphology on a continuous spectrum, and risk stratification and the approach to repair depend on patient-specific BAV morphology \cite{michelena_international_2021}. Moreover, the dynamic features of the valve - such as prolapse and coaptation - are critical information for repair planning. Currently, pre-operative BAV evaluation relies on transesophageal echocardiography (TEE) and intra-operative observation (direct visualization of the valve when the heart is arrested on cardiopulmonary bypass). However, the quality of TEE is variable and can be significantly impacted by noise. A drawback of intra-operative visualization is that the valve is in a flaccid, unpressurized state while the heart is arrested. 4D computed tomography (CT) is an alternative modality that produces volumetric scans of the aortic valve  over the cardiac cycle. Already routinely used to plan transcatheter aortic valve replacement (TAVR) \cite{francone_ct_2020}, 4D CT may also be a valuable tool for assessing BAV morphology and dynamics for surgical repair planning. 
\begin{figure}[htb]
    \centering
    \includegraphics[width=0.7\linewidth]{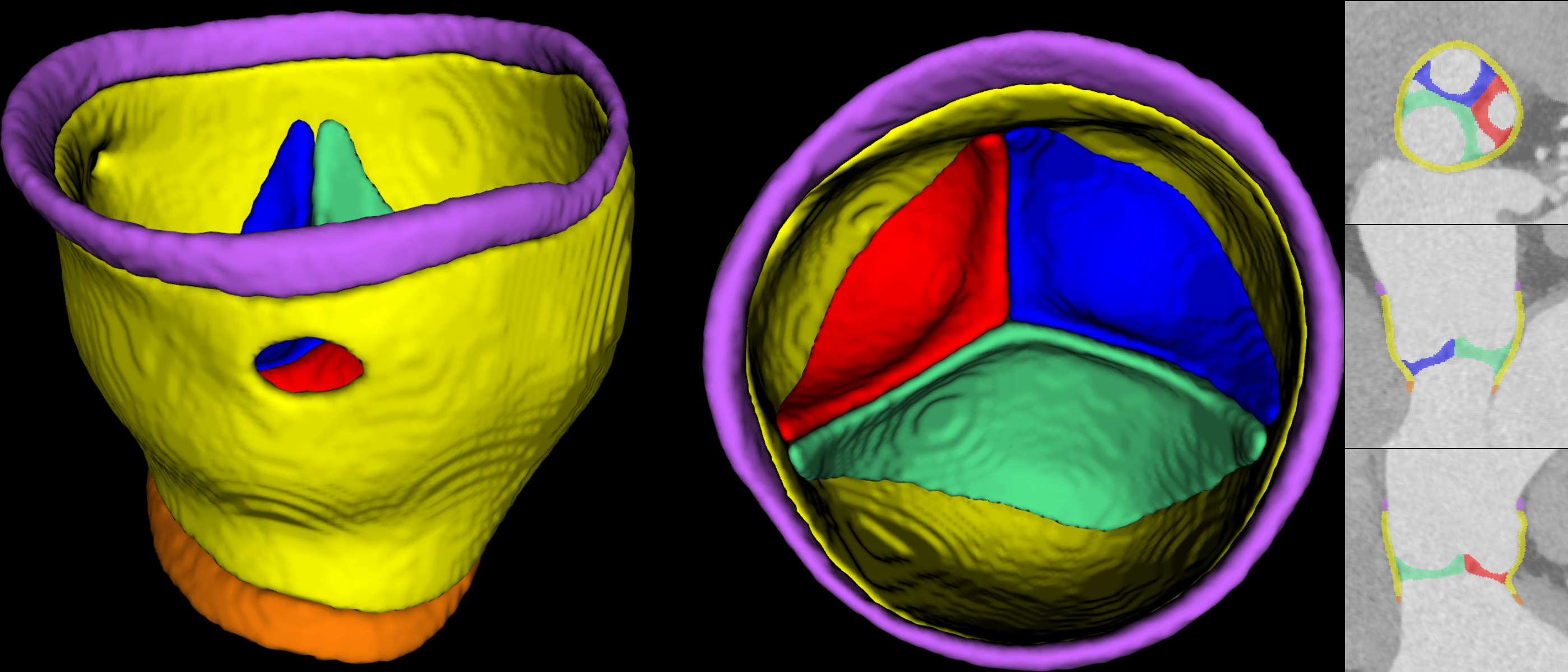}
    \caption{Anatomy of a BAV with left-right coronary cusp fusion. 3D side view (left), aortic view (center) and 2D views (right) are shown. The structures segmented are the left coronary cusp (LCusp, red), non-coronary cusp (NCusp, green), right coronary cusp (RCusp, blue), root wall (yellow), left ventricular outlet (LVO, orange) and sinotubular junction (STJ, purple)}
    \label{fig:anatomy}
\end{figure}

Segmentation of aortic cusps and root wall is a crucial step in creating patient-specific BAV models. Quantitative analysis in 4D CT by Fikani et al. \cite{fikani_morphological_2024} and Ionasec et al. \cite{ionasec_dynamic_2008}  demonstrated the utility of temporal data in generating patient-specific models and quantifications, but only on normal tricuspid aortic valves. The last decade has seen the rise of semi- and fully automated methods for heart valve segmentation, many of which were developed for TEE of the mitral \cite{chen_automatic_2023} \cite{carnahan_deepmitral_2021}, aortic \cite{pouch_segmentation_2015}, and tricuspid \cite{herz_segmentation_2021} valves. Pak et al. \cite{pak_efficient_2020} trained a Spatial Transformation Network to perform multi-class 3D CT segmentation of the aortic valve; however, the study did not include congenital BAVs. Many of these examples leverage supervised deep learning due to its fast inference time, good accuracy, and capacity for full automation \cite{singh_3d_2020}. However, many studies in the area of image segmentation do not convey the clinical usability of the segmentations and only report global metrics such as Dice scores. In the case of BAV repair planning, clinical quantification of the valve morphology is crucial in determining the feasibility and specific approaches to repair \cite{ehrlich_state--art_2020} \cite{isselbacher_2022_2022}. Thus, the main objective of this work is to demonstrate fully automated 4D CT segmentation that produces clinically informative BAV measurements. 

In this study, we generate fully labeled 4D CT data of minimally calcified BAVs and train a neural network to segment the aortic cusps and root wall, with demarcation of the sinotubular junction and left ventricular outlet. In addition to evaluating the accuracy of fully automated segmentation, we leverage the output to measure geometric cusp height, commissural angle configuration, and annular diameter, which are crucial in BAV surgical planning and demonstrate the translational potential of the segmentation pipeline. Our main contributions are: 1) fully automated multi-class segmentation of 4D CT images of minimally calcified BAVs, 2) evaluation of the temporal consistency of fully automated segmentation, and 3) assessment of the efficacy of automated segmentation for performing surgically relevant parameter measurements. 

\section{Method}
\subsection{Data Collection}
The study enrolled adults with minimally calcified BAVs \footnote{This research study was approved by the Institutional Review Board of University of Pennsylvania.}. A total of 11 scans were acquired from 10 patients, yielding a total of 188 individual 3D frames.  One patient discovered to have a trileaflet aortic valve was excluded from the study. 4D contrast-enhanced CT images of the patients' aortic valves were acquired with a modified TAVR acquisition protocol, including at least one complete cardiac cycle consisting of 10 to 20 frames and a voxel size range of $(0.367 \sim 0.625) \times (0.367 \sim 0.625) \times (0.2 \sim 0.4) \, \text{mm}^{3}$. During acquisition, the aortic leaflets, root wall, sinotubular junction and left ventricular outlet were included in the field of view. Images were exported to DICOM files and de-identified. 
\subsection{Generation of 4D Ground Truth Segmentations}
The creation of ground truth 4D segmentations began with identification of cardiac phases in each 4D series and manual segmentation of two reference frames. For each 4D series, each 3D frame was manually inspected in ITK-SNAP \cite{yushkevich_user-guided_2006} and classified as either a diastolic or systolic frame, and one diastolic frame and one systolic frame were selected as references. The following six structures were manually segmented as shown in Figure \ref{fig:anatomy}: left coronary cusp, non-coronary cusp, right coronary cusp, the root wall, left ventricular outlet  and sinotubular junction. Manual segmentations were carried out by an expert using ITK-SNAP and were reviewed by at least one other experienced member of the group. 
\begin{figure}[htb]

\begin{minipage}[b]{0.85\linewidth}
  \centering
  \centerline{\includegraphics[width=8.5cm]{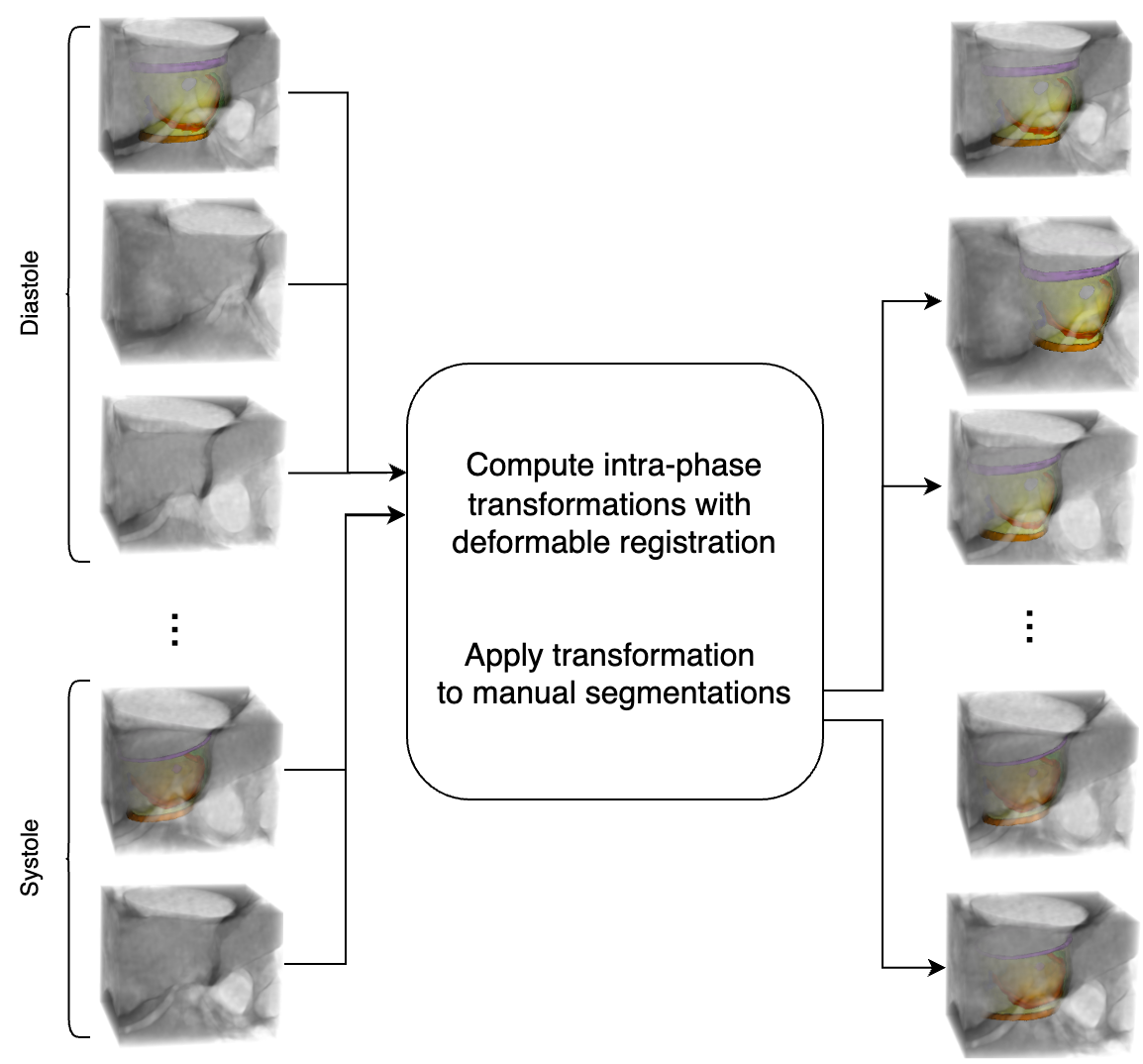}}
\end{minipage}

\caption{Illustration of the semi-automatic ground truth segmentation workflow. For each 4D scan, one 3D diastolic frame and one systolic frame are manually annotated. Within each cardiac phase, deformable transformations are calculated from the manually labeled frame to each of the remaining frames, and the transformations are applied to the manual segmentation to obtain segmentations of all remaining frames.}
\label{fig:res}
\end{figure}
The reference diastolic and systolic frames were then propagated to all diastolic frames and systolic frames, respectively. Deformable transformations from each of the non-reference frames to the reference frames were calculated using an extension of the Greedy algorithm \cite{yushkevich_icp174_2016} as described in \cite{aggarwal_strain_2023}. The transformations were then inversely applied to the reference segmentations in order to semi-automatically generate segmentations for all non-reference frames. 
\subsection{Fully Automated Image Segmentation}
We trained a series of 3D full-resolution nnU-Net \cite{isensee_nnu-net_2021} models to perform fully automated segmentation of the six label classes. nnU-Net is a self-configuring convolutional neural network architecture that has established itself as a good baseline for deep learning-based segmentation models [10]. Because nnU-Net does not support 4D segmentation, 4D series were sliced into 3D image volumes that were segmented independently. Due to the small number of patient data available, we employed a leave-one-out cross validation (LOOCV) to evaluate the automated segmentations. We employed a nested cross-validation method: for each nnU-Net trained, one patient's 4D image was held out for testing, while the remaining patients' images were used to perform an inner 5-fold cross-validation. We trained networks in full-resolution 3D (3d\_fullres) mode for 250 epochs with stochastic gradient descent. All other hyperparameters were kept at the nnU-Net default settings. At inference, the test set was processed by all five folds and voxel classification was performed by majority voting between the five folds. 
\subsection{Metrics for Segmentation Evaluation}
\begin{figure}[htb]
    \centering
    \includegraphics[width=0.75\linewidth]{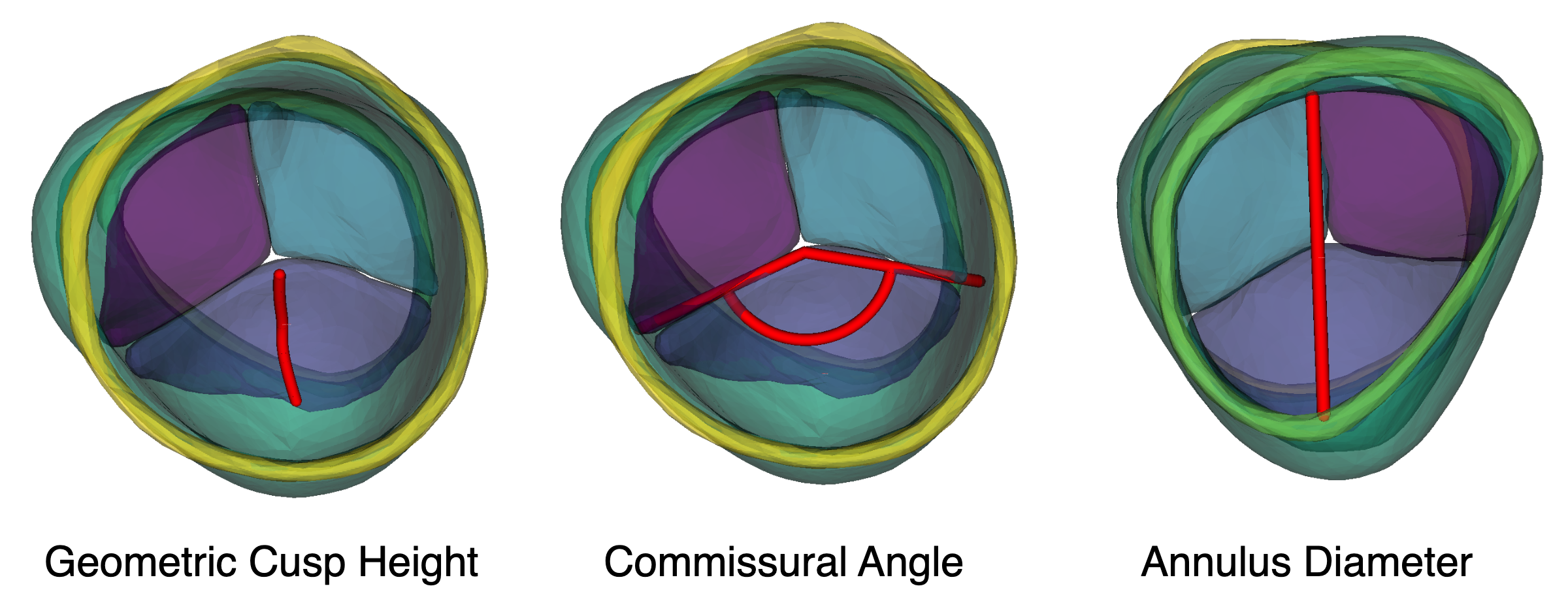}
    \caption{Illustration of measurement protocols using the Markup module in 3D Slicer. The left two panels show the valve from an aortic perspective, and the right panel from a ventricular perspective. }
    \label{fig:3d_slicer}
\end{figure}
We evaluated all automated segmentations against the ground truth with respect to four categories of metrics: Dice similarly, symmetric mesh distance \cite{pouch_medially_2015}, aortic outflow orientation accuracy, and clinical measurements. Dice scores were calculated with SimpleITK, and symmetric mesh distances were calculated using the cm-rep meshdiff executable \cite{yushkevich_continuous_2009}. Accuracy of orientation of the predicted segmentation was evaluated to ensure predicted segmentions align with the ground truth regarding STJ and LVO, as this is crucial for surgical planning. We calculated the angle between vectors spanning the LVO to STJ in the predicted and ground truth segmentations; an angle between 0 to 90 degrees indicated that the predicted segmentation had the same orientation as the ground truth, whereas an angle outside that range indicated the orientation was flipped. Lastly, for the clinical measurements, three experienced observers measured geometric cusp height of the non-fused cusp, annulus diameter, and commissural angle of all predicted and ground truth segmentations using the 3D Slicer Markup module \cite{kikinis_3d_2014} as shown in Figure \ref{fig:3d_slicer}. Specifically, geometric cusp height was measured on the non-fused cusp from the center of the nadir to the center of the free margin. Annulus diameter was measured from the nadir of the non-fused cusp to the opposite side of the aortic root wall. Commissural angle (symmetry of the BAV) was measured in a plane parallel to a projection of the annulus.

\section{Results}
\begin{figure}[htb]
  \begin{minipage}[b]{.48\linewidth}
    \centering
    \centerline{\includegraphics[width=4.0cm]{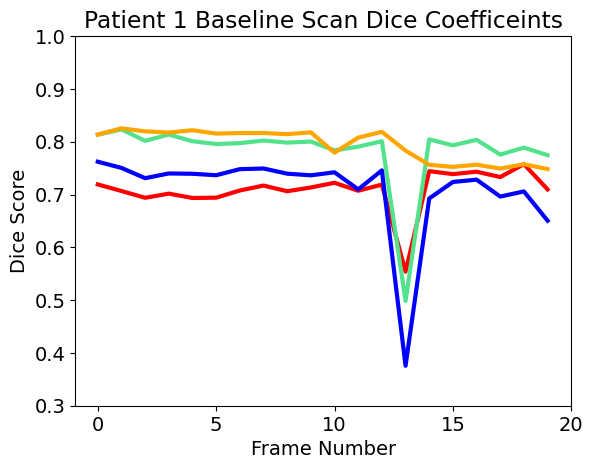}}
  \end{minipage}
  \hfill
  \begin{minipage}[b]{0.48\linewidth}
    \centering
    \centerline{\includegraphics[width=4.0cm]{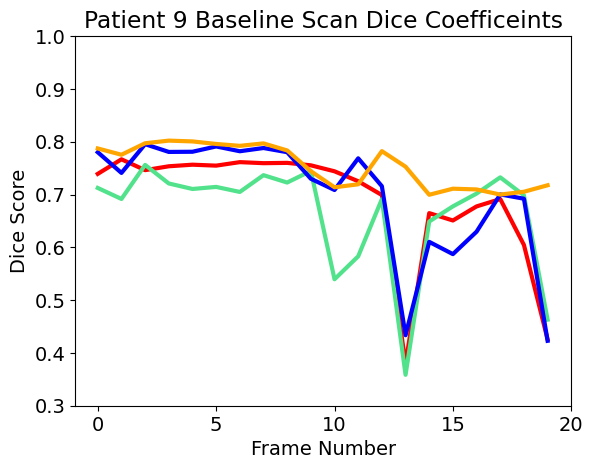}}
  \end{minipage}
  
  \caption{Plots of Dice scores of LCusp (red), NCusp (green), RCusp (blue), root wall (yellow) as a function of frame number. Segmentation results with strong (left) and weaker (right) temporal consistency are shown. The significant drop in Dice scores on the left subplot demonstrates that the deep learning model struggles to accurately segment the transitional frame during a cardiac cycle when the valves are in the process of opening.}
  \label{fig:dice_plot}
\end{figure}
\begin{figure}[htb]
    \centering
    \includegraphics[width=\linewidth]{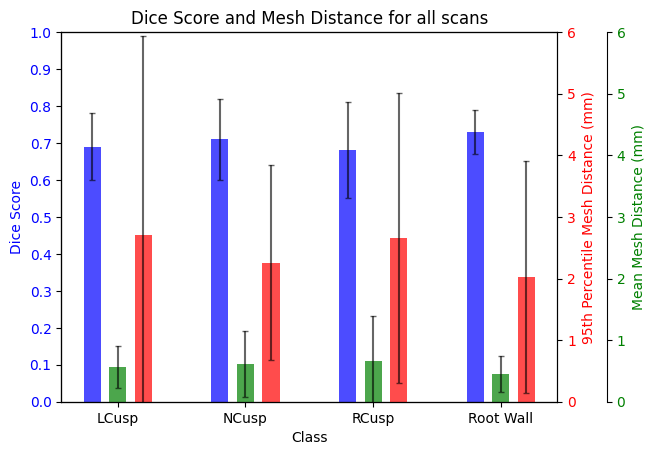}
    \caption{Segmentation accuracy metrics: Dice, mean symmetric mesh distance, and 95th percentile distance for all three cusp components and the aortic root wall. The metrics compare the predicted and ground truth segmentations across all test sets. }
    \label{fig:dice_and_meshdist}
\end{figure}

An example of a predicted segmentation by nnU-Net is shown in Figure \ref{fig:anatomy}. Overall, the models achieved a Dice score of $0.69 \pm 0.09$ for the LCusp, $0.71 \pm 0.11$ for the NCusp, $0.68 \pm 0.13$ for the RCusp and $0.73 \pm 0.06$ for root wall. Figure \ref{fig:dice_plot} shows the Dice scores plotted against frame number for a temporally consistent segmentation performance (left) and a segmentation that is weaker in temporal consistency (right). The average and 95th percentile mesh distances between the predicted and ground truth segmentations are $0.57\pm0.34$mm and $2.70\pm3.24$mm for LCusp, $0.61\pm0.53$mm and $2.26\pm1.58$mm for NCusp, $0.66\pm0.73$mm and $2.66\pm2.35$mm for RCusp and $0.45\pm0.29$mm and $2.02\pm1.88$mm for root wall. The average Dice scores for each test 4D scan are shown in Figure  \ref{fig:dice_and_meshdist}. Note that we present the Dice and mesh distances for four out of six labels (three cusps and root wall) since the remaining two labels (STJ and LVO) are demarcations of orientation, and their accuracy is reflected by the aortic outflow orientation calculated in offset angles. The offset angles between the predicted and ground truth segmentations have an average of $9.26 \pm 6.0$ degrees and range of $[0.17, 24.74]$ degrees. None of the offset angles exceeded 90 degrees, indicating that all predicted segmentations had the correct aortic outflow orientation. 
The difference in geometric cusp height, commissural angle and annulus diameters measurements in the ground truth and predicted segmentations are shown in Table \ref{tab:measure_mean_diff} and Table \ref{tab:measure_max_diff}. Table \ref{tab:measure_max_diff} shows the result of intraclass correlation analysis (ICC), which quantifies the inter-rater consistency in each measurement. 

\begin{table}
    \centering
    \resizebox{0.48\textwidth}{!}{%
    \begin{tabular}{|c|c|c|c|} \hline 
         &  \textbf{Geometric Cusp Height (mm)} &  \textbf{Commissural Angle (degrees)} & \textbf{Annulus Diameter (mm)} \\ \hline 
         \textbf{Rater 1} & \makecell{2.07 ± 1.63 \\ (p=0.1)} & \makecell{5.9 ± 4.86 \\ (p=0.91)} & \makecell{1.17 ± 0.77 \\ (p=0.93)} \\ \hline 
         \textbf{Rater 2} & \makecell{1.90 ± 1.26 \\ (p=0.23)} & \makecell{6.9 ± 5.69 \\ (p=0.82)} & \makecell{1.00 ± 0.52 \\ (p=0.91)} \\ \hline 
         \textbf{Rater 3} & \makecell{1.82 ± 1.74 \\ (p=0.19)} & \makecell{8.09 ± 6.82 \\ (p=0.65)} & \makecell{1.32 ± 1.24 \\ (p=0.77)} \\ \hline
    \end{tabular}%
    }
    \caption{Average absolute difference between measurements derived from the ground truth and predicted segmentations, with the p-values of paired Student's t-tests. }
    \label{tab:measure_mean_diff}
\end{table}

\begin{table}
    \centering
    \resizebox{0.48\textwidth}{!}{%
    \begin{tabular}{|c|c|c|c|} \hline 
         &  \textbf{Geometric Cusp Height (mm)}&  \textbf{Commissural Angle (degrees)}& \textbf{Annulus Diameter (mm)}\\ \hline 
         \textbf{Ground Truth}&  \makecell{1.45 ± 1.32 \\ (ICC=0.81)}&  \makecell{17.59 ± 8.45 \\ (ICC=0.75)}& \makecell{1.70 ± 1.52 \\ (ICC=0.94)}\\ \hline 
         \textbf{Predicted}&  \makecell{1.64 ± 0.56 \\ (ICC=0.80)}&  \makecell{17.15 ± 4.67 \\ (ICC=0.58)}& \makecell{1.12 ± 0.97 \\ (ICC=0.97)}\\ \hline
    \end{tabular}%
    }
    \caption{Average of maximum absolute difference among three raters for each segmentation, with the corresponding ICC among three raters. }
    \label{tab:measure_max_diff}
\end{table}

\section{Discussion}
This study is, to our knowledge, the first that performs multi-class CT segmentation of the bicuspid aortic valve across the full cardiac cycle. BAV repair surgery has gained traction in recent years  \cite{ehrlich_state--art_2020}, but involves complex decision making related to valve repair suitability and the surgical approach. The diversity of BAV morphology necessitates careful surgical planning informed by precise and reliable patient-specific BAV modeling. Towards this end, 4D CT can provide high quality images of the valve apparatus. In addition to evaluating automated BAV segmentation performance in 4D CT with respect to the conventional Dice coefficient, we incorporated measurements that are directly related to BAV treatment planning in order to assess its translational potential. 

The Dice scores achieved in this study (around 0.7) are consistent with previous studies of automated heart valve leaflet segmentation \cite{aly_fully_2022}, which typically have Dice scores below 0.9 since leaflets are thin sheet-like structures. In most cases as shown in Figure \ref{fig:dice_and_meshdist}, the average symmetric mesh distances between automated and ground truth segmentations are within two voxels, and the 95th percentile distances are between 4 to 7 voxels. 

To evaluate automated segmentation from a clinical perspective, we incorporated three measurements that are used to guide risk stratification and the approach to BAV repair surgery. As shown in Table \ref{tab:measure_mean_diff}, for annular diameters, the max differences are smaller than the difference in thresholds for determining repair eligibility in a guideline document by the American Heart Association \cite{isselbacher_2022_2022}. Studies have also shown that commissural angle and geometric cusp height, which is a measure of the amount of available cusp tissue for repair, are also important in determining the outcomes of BAV repair operations \cite{ehrlich_state--art_2020} \cite{de_kerchove_variability_2019}. We also calculated the ICC among three raters for each of the measurements. Commissural angle has a relatively lower ICC primarily because the inter-observer subjectivity in defining the annular plane that serves as a reference for the angle measurement; however, an ICC of 0.58 on predicted segmentations and 0.75 on ground truth segmentations still demonstrate moderate consistency among raters. The consistency for geometric cusp height and annulus diameter are excellent as shown by ICC scores above 0.8 in Table \ref{tab:measure_max_diff}. The high ICC further demonstrates that accurate measurements can be obtained from fully automated segmentations. 

This study has several limitations. With only 11 4D scans totaling 188 individual 3D CT images, segmentation performance could benefit from increasing training data to capture more diverse aortic valve morphologies, including trileaflet aortic valves. Another limitation is that nnU-Net is designed for 3D segmentation rather than analysis of 4D image series, so the network was trained with individual 3D frames without accounting for temporal consistency between frames over the cardiac cycle. As shown in Figure \ref{fig:dice_plot}, the Dice scores vary over the cardiac cycle, indicating that the results produced by nnU-Net have limited temporal consistency. This method could benefit from future adaptations that enforce temporal constraints on volumes within the same image series. 
\section{Conclusion}
This study proposed a fully automated segmentation pipeline for 4D CT of bicuspid aortic valve based on the nnU-Net architecture, and evaluated its performance on both global and clinically relevant metrics. Trained models can achieve a segmentation accuracy on par with previous heart valve segmentation literature, and the predicted segmentations can be used to generate consistent clinical measurements of the aortic valve when compared to manual segmentation. Future work may focus on taking full advantage of 4D data by enforcing temporal consistency in volumetric time series. 

\section{Acknowledgments}
This study was supported by the National Institutes of Health (R01-HL163202), the Translational Bio-Imaging Center at the University of Pennsylvania, the Children's Hospital of Philadelphia (CHOP) Innovation Award, and the Topolewski Pediatric Valve Center at CHOP.

\begingroup
\setstretch{0.8}  
\endgroup

\setlength{\itemsep}{0.3em}
\bibliographystyle{IEEEtran}
\def\url#1{}
\bibliography{isbi, quals}

\end{document}